 \documentclass[showpacs,showkeys,preprintnumbers,amsmath,amssymb,aps]{revtex4}
\usepackage{graphicx}
\usepackage{dcolumn}
\usepackage{bm,color}
\usepackage{epsfig}

\newcommand{\ede}{\end{document}}  

\def\be{\begin{eqnarray} &&}
 
 \def\ee{\end{eqnarray}}
\def\psla{\rlap \slash} 

\begin{document}
 
\title{ 
Pion Electromagnetic Form Factor at Lower and Higher Momentum Transfer 
}
\author{J.~P.~B.~C.~de~Melo,~R\^omulo Moreira Moita and Kazuo Tsushima}
\affiliation{ Laborat\'orio de 
F\'{\i}sica Te\'orica e Computacional~-~LFTC \\ 
 Universidade Cruzeiro do Sul, 01506-000, S\~ao Paulo, Brazil}
\date{\today}

\begin{abstract}
The pion electromagnetic form factor is calculated 
at lower and higher momentum transfer in 
order to explore constituent quark models and the differences 
among those  models. In particular, the light-front constituent quark model is utilized here to 
calculate the pion electromagnetic form factor at lower and 
higher energies. The matrix elements of the 
electromagnetic current, are calculated with 
both "plus" and "minus" components of the electromagnetic current in the light-front. 
Further, the electromagnetic form factor is compared with 
other models in the literature and experimental data. 
\end{abstract}
\date{\today} 
\maketitle

\section{Introduction}   

Quantum cromodynamics~(QCD), is believed as the correct 
theory of the strong interactions which is 
one of the four fundamental interactions of nature. 
One of the most important questions in QCD not 
yet solved, is the  
\linebreak non-perturbative regime. However, with the relativistic 
constituent quark model~(CQM), is possible to obtain answers for hadronic physics 
in terms of the degrees freedom from the QCD, 
i.e., quarks and gluons~\cite{Brodsky98}. 
Before the advent of QCD, the pion, the lowest mass hadronic bound state,  
providing the long-range attraction part of the nucleon-nucleon 
potential~\cite{Lacombe2002} with the intermediate energies. 
The main proposal of the light-front approach 
is to describe consistently the hadronic pion bound state to both, higher and lower 
momentum transfer regime.  

With this purpose, the light-front quantization is utilized to 
compute the hadronic bound state wave functions~\cite{Brodsky98,Harina96},
which are simpler compared 
to the instant form quantum field theory~\cite{Zuber}. 

As known, in the light-front models, the bound states wave functions are
defined in the hypersurface 
\linebreak $x^+=x^0+x^3=0$, and the wave functions are covariant under
kinematical front-form boosts, because the Fock-state decomposition stability~\cite{Perry90}. 
The bound state wave functions with the light-front constituent quark model 
(LFCQM) have received much attention lately~\cite{Terentev76,Wilson95}.

The quark models show an impressive success in the description of the 
electromagnetic properties of the hadronic wave functions,
for pseudoscalar and spin half particles~\cite{Dziembowski87,Cardarelli95,
Cardarelli96,Pacheco99,Simula2002,Pacheco2004,Hwang2003,Huang2004,Braguta2004,Salcedo2004,
Salcedo2006,Karmanov2007,Pacheco2002,Ji2001,Choi2001,Tsirova2009,Melo2006,Pena2014,Yabusaki2015,Horn2016,Vary2016} 
and also vector particles~\cite{Pacheco97,Pacheco992,Lev1999,Lev2000,Jaus2003,Aliev04}.

The extraction of the electromagnetic form factor with the light-front approach 
depends on which component of the electromagnetic 
current is utilized to calculate the form-factors, 
due to the problems related with the rotational 
symmetry breaking and the zero modes, 
namely a non-valence 
contribuition to the matrix elements of the 
electromagnetic current~\cite{Pacheco97,Pacheco98,Naus98,Bakker2003,Choi2004,Otoniel12}.

It is found in references~\cite{Pacheco97,Pacheco992,Bakker2002,Bakker2003,Choi2004,Pacheco2012,Clayton2015} 
for spin-1 particles, that the plus component
of the electromagnetic current,~("$J^{+}$"), is not free from the
pair terms contribution~(or non-valence contributions) in the Breit frame 
($q^+=0$), and thus the rotational symmetry is broken. 

Then, the matrix elements of the electromagnetic 
current in the light-front formalism
have other contributions for the 
electromagnetic current besides the valence contribution; 
that contribution corresponds to the non-valence components or pair terms
added to the matrix elements of the
electromagnetic current~\cite{Pacheco2002,Pacheco98,Naus98,Pacheco2012} 
in order to restore the covariance.

If the pair terms contribution is taken correctly, it doesn't matter
which component of the electromagnetic current is utilized 
in the light-front formalism in order to 
extract the electromagnetic form factors of the hadronic bound states. 
In the present work, two types of the vertex functions
are utilized in order to calculate the pion
electromagnetic form-factor for the 
$\pi-q\bar{q}$ vertex 
and both are compared with the 
experimental 
data~\cite{Amendolia84,Amendolia86,Frascati2001,Volmer2001,Blok2002,Horn06,Tadevosyan2006}. 

At lower momentum transfer, non-perturbative regime 
of QCD is more important compared 
with the higher momentum transfer of the perturbative regime 
of QCD. Perturbative QCD works well 
over the momentum transfer squared $1.0$~(GeV/c)$^2$ and is predominant around~$5.0$~(GeV/c)$^2$. 
The studies on light vector and scalar mesons are important, 
because they give a direction for understanding 
why QCD works in the non-perturbative regime and also, 
the light mesons  are related with the chiral symmetry breaking. 

Hadronic bound states,~mesons, are described with other approaches 
in references~\cite{Roberts96,Hawes99,Maris2000,Krutov2001,Aliev04,
Carvalho2004,Desplanques2004,Desplanques2009,Noguera07,Pinto05,
Egle2005,Nesterenko82,Rady01,Braguta041,Braguta042}.
~Further, another possibility is to study the hadronic bound state 
with the lattice formulation in the light-front~\cite{Dalley01}.

For the lightest pseudoscalar bound state meson, the models with the Schwinger-Dyson  
equations~\cite{Roberts96,Hawes99,Maris2000} describe the 
electromagnetic form factor quiet well, 
however some differences among the models can be noticed in the literature. 

Here, the light-front models for the pion which were presented in previous 
work~\cite{Pacheco99,Pacheco2002},  are extended to higher 
momentum transfer and compared with other 
quark models, for example, the vector meson dominance~\cite{Kroll67,Krein93}.

This paper is presented in the following; in section II, the model of the wave function for the
bound quark-antiquark in the light-front is presented,
and the electromagnetic form factor 
is calculated with non-symmetric and symmetric vertex $\pi-q\bar{q}$, 
also, in the case of non-symmetric vertex, the plus and minus 
components of the electromagnetic current are used. As well, in this section, 
it is presented the calculation for the weak decay constant for the pion.

In section III, the vector dominance model is presented in order 
to compare with the light-front approach utilized here. 
Finally, in  section IV, the numerical results and discussions are given, and 
the conclusions are presented in section V.

\section{Light-Front Wave Function and Electromagnetic Form Factor}

In the light-front formalism, the main goal is to solve the bound state problem, 
that is translated in solving the equation below, 
\begin{equation}
H_{LF}| \Psi > = M^2| \Psi> \ .
\label{auto}
\end{equation} 
In Eq.~(\ref{auto}), 
the light-front Hamiltonian $H_{LF}$ has the eigenvalues given by the invariant mass $M^2$,  
where the eigenvalues are associated with the physical particles, the eigenstates of the 
light-front Hamiltonian~\cite{Brodsky98}. 
The hadronic light-front wave function is related with 
Bethe-Salpeter wave function~(see Ref.~\cite{Pacheco2002} for more details about this point). 
With the light-front wave function it is possible to calculate the matrix elements 
between hadronic bound states. In the light-front, the meson bound state 
wave function is a superposition of all Fock states, and the wave function is given by
\begin{equation}
|\Psi_{meson}>=
\Psi_{q\bar{q}} |q\bar{q}>  +
\Psi_{q\bar{q}g}|q\bar{q}g>  +  \cdots .
\label{eq2}
\end{equation}

With the light-front hadronic wave function above, it is possible to 
calculate the hadronic electromagnetic 
form factors from the overlap of light-front wave functions between the 
final and initial states. 
In general, the electromagnetic form-factor for the pion 
is expressed by the covariant 
equation below, 
\begin{equation}
(p+p^{\prime})^{\mu} 
F_{\pi}(q^2) \ = \ <\pi(p^{\prime})|J^{\mu}|\pi(p)>, 
\ \ \ \      q=p^\prime -p \ ,
\label{ffactor}
\end{equation}
where $J^{\mu}$ is the electromagnetic current, which is possible to be expressed
in terms of the quark fields~$q_f$~and the charge 
$e$~($f$ is the flavor of the quark field):
$J^{\mu}=\sum_{f} e_{f} \bar{q}_{f} 
\gamma_\mu q_{f}$. 
The matrix elements of the electromagnetic current, 
are written according to the following equation:   
\begin{equation}
J^\mu =-\imath 2 e \frac{m^2}{f^2_\pi}
N_c\int \frac{d^4k}{(2\pi)^4} Tr \Bigl[ S(k)
\gamma^5 S(k-p^{\prime})
\gamma^\mu S(k-p) \gamma^5 \Bigr]
\Gamma(k,p^{\prime})
\Gamma(k,p) \ , 
\label{jmu}
\end{equation}
where 
$\displaystyle S(p)=\frac{1}{\rlap\slash p-m+\imath \epsilon}$ 
is the quark propagator 
and $N_c=3$ is the number of colors, and m is the constituent quark mass. 
The calculation here is made in 
the Breit frame, 
with $p^{\mu}=(0,-q/2,0,0)$ and~$p^{\prime {\mu}}=(0,q/2,0,0)$  
for the initial and 
final momenta of the system respectively, and 
the momentum transfered is 
$q^{\mu}=(0,q,0,0)$ and $k^{\mu}$
is the spectator quark momentum. The 
factor 2 appears from the isospin algebra~\cite{Pacheco99,Pacheco2002}. 
The function $\Gamma(k,p)$, is the regulator vertex function 
used in the present work to regularize the Feynman amplitude, ie., 
the triangle diagram for the electromagnetic 
current,~Eq.~(\ref{jmu}), written above.

Here, we have utilized two possible~$\pi-q\bar{q}$ vertex functions; the first one is the
non-symmetric vertex, used in the previous work~\cite{Pacheco99,Otoniel12}:
\begin{equation}
\Gamma^{(NSY)}(k,p)=
\biggl[
\frac{N}{((p-k)^2-m^2_R+\imath\epsilon)}
\biggr]
\label{nosymm} \ ,
\end{equation} 
and the second one is, a symmetric vertex, used in the 
references~\cite{Pacheco2002,Yabusaki2015}:
\begin{equation}  
\Gamma^{(SY)}(k,p)=
\biggl[ 
\frac{N}{(k^2-m^2_R + \imath\epsilon)} +
\frac{N}{((p-k)^2-m^2_R + \imath\epsilon)}
\biggr].
\label{symm}
\end{equation} 
In the expressions for the vertex above, $m_R$ is the regulator mass used in order to keep 
the amplitudes finite and, also, represents the soft effects at the short range. 
An important question in QCD  and electromagnetic processes is the 
current conservation, which is a consequences of the gauge invariance. 
It is also important to check with the vertex functions, $\Gamma(k,p)$, 
utilized in the present work, 
the electromagnetic current conservation. 
The current conservation 
is easily proved in the Breit frame~(see the reference~\cite{Naus98}, for this point).

The $J^{+}$~component of the electromagnetic current is used to extract the pion
electromagnetic form factor from Eq.~(\ref{ffactor}), where the Dirac "plus" matrix is given by   
$\gamma^+=\gamma^0+\gamma^3$. The plus component, $J^{+}_{\pi}~(=J^0+J^3)$, of the electromagnetic current 
for the pion calculated in the light-front formalism through the triangle 
Feynman diagram in the impulse approximation, which represents the 
photon absorption process by the hadronic bound state of  
the $q\bar{q}$ pair, is given by: 
\begin{eqnarray}
J^+_\pi& = & 2  e (p^{+}+p^{\prime +}) F_\pi(q^2) \nonumber \\
& = & \imath e
\frac{m^2}{f^2_\pi} 
N_c \int 
\frac{dk^-dk^+d^2k_{\perp}} {2 (2\pi)^4}
\frac{ Tr[ {\cal O}^{+} ]\Gamma(k,p^{\prime})
\Gamma(k,p)} 
{k^+ (k^- - \frac{f_1-\imath \epsilon}{k^+})} \nonumber   \\ 
&   & 
\times 
\biggl[ 
\frac{1}{(p^+ - k^+)(p^{-}-k^- - 
\frac{f_2 -\imath \epsilon }{p^{+} - k^+})  
(p^{\prime+} - k^+)(p^{\prime-}-k^- - 
\frac{f_3-\imath \epsilon }
{p^{\prime+} - k^+}) } 
\biggr], 
\label{jpion}
\end{eqnarray}
where the $f_i~(i=1,2,3)$ functions above, are defined 
by,~$f_1=k_{\perp}^2+m^2$,~$f_2=(p-k)_{\perp}^2+m^2$ and 
$f_3=(p^{\prime}-k)_{\perp}^2+m^2$,~with the light-front coordinates 
defined,  $a^{\pm}=a^0 \pm a^3$ and 
$\vec{a}_{\perp}=(a_x,a_y)$~\cite{Brodsky98,Harina96}.

In the expression of the electromagnetic current,~Eq.(\ref{jpion}), the 
Jacobian for the transformation to the light-front coordinates is 
$1/2$, and the Dirac trace in the Eq.~(\ref{jpion}) for 
the operator $\cal{O}^{+}$ is written in the light-front coordinates in 
the Breit frame with Drell-Yan condition~($q^+=0$),~as:
\begin{eqnarray}
& & Tr[ {\cal O }^{+}  ]= Tr[ (\psla{k}+m)
\gamma^5 
(\psla{k}-\psla{p^{\prime}}+m)
\gamma^{+}
(\psla{k}-\psla{p}+m)  
\gamma^5 ] \nonumber \\ 
&  &  = [-4 k^- (k^+-p^+)^2 + 4 (k^2_\perp+m^2) (k^+-2
p^+) + k^+ q^2].
\end{eqnarray}
The quadri-momentum integration of the Eq.~(\ref{jpion}) has two contribution intervals: 
\newline 
(i) $0<k^+<p^+$ and (ii) 
$p^+<k^+<p^{\prime +}$, where 
$p^{\prime +}=p^+ + \delta^+$.
\par
The first interval,~(i), is the contribuition to the valence wave function for 
the electromagnetic form factor, and the second,~(ii), corresponds to the pair terms 
contribution to the matrix elements of the electromagnetic current. In the case of the
non-symmetric vertex with the plus component of the electromagnetic current, 
the second interval does not give any contribution for the current matrix elements, 
because the non-valence terms contribution in this case is zero~\cite{Pacheco99,Otoniel12}.  

However, it is not the case for the minus component of the electromagnetic current for the pion, 
where beyond the valence contribution, we have a non-valence 
contribution~\cite{Pacheco99} for the matrix 
elements of the electromagnetic current.  For the first interval integration, 
the pole contribution is 
$\bar{k}^-=\frac{f_1-\imath \epsilon}{k^+}$. 
After the integration for the light-front energy,~$k^-$, the
 electromagnetic form factors with nons-symmetric 
vertex  and the plus component of the
 electromagnetic current is given by 
\begin{eqnarray}
F^{+(i){(NSY)}}_{\pi}(q^2) 
=  & &  2 \imath e 
\frac{m^2 N^2}{ 2 p^+ f^2_\pi} N_c \int \frac{%
d^{2} k_{\perp} d k^{+}}{2(2 \pi)^4} 
\biggl[
\frac{ Tr [ {\cal O}^{+} ] } {k^+(p^{+} - 
k^+)^2
(p^{+}-k^+)^2} 
 \nonumber  \\  & & \times 
\frac{~\theta(k^+) \theta(P^+ - k^+)} 
{(p^- - \bar{k}^- - \frac{f_2 -\imath \epsilon }{p^+ - k^+})
(p^{-} - \bar{k}^- - \frac{f_3 -\imath \epsilon }{p^{+} - k^+}) 
} 
  \nonumber \\   & & \times 
\frac{1} {
(p^- - \bar{k}^- - \frac{f_4 -\imath \epsilon }
{p^+ - k^+}) (P^{'-} - \bar{k}^- - 
\frac{f_5 -\imath \epsilon }{p^{'+} - k^+})}
\biggr], 
\label{eq5.6}
\end{eqnarray}
where the Dirac trace above (for the quark 
on-shell) is:
\begin{eqnarray*}
& & Tr[ {\cal \bar{O} }^{+} ]=   
[-4 \bar{k}^- (k^+-p^+)^2 + 4 (k^2_\perp+m^2) (k^+-2
p^+) + k^+ q^2].
\end{eqnarray*}
The functions 
$f_1$,~$f_2$ and $f_3$ were already defined and 
the new functions above are,
~$f_4=(p-k)_{\perp}^2+m^2_R$ and
$f_5=(p^{\prime}-k)_{\perp}^2+m^2_R$. 
The light-front wave function for the pion with the non-symmetric vertex is 
\begin{equation}
\Psi^{(NSY)}(x,k_{\perp})=
\biggl[
\frac{N}{(1-x)^2 
(m_{\pi}^2-{\cal M}_0^2) (m_{\pi}^2-{\cal M}_R^2)}
\biggr] ,
\label{wavefunction}
\end{equation}
where the fraction of the carried momentum by the quark 
is $x=k^{+}/p^{+}$ and ${\cal M}_R$ function is written as 
\begin{equation}
{\cal M}_R^2={\cal M}^2(m^2,m^2_R)=
\frac{k_{\perp}^2+m^2}{x}+
\frac{(p-k)_{\perp}^2+m_R^2}{(1-x)}-p^2_{\perp} \ .
\end{equation}
In the pion wave function expression, ${\cal M}^2_0={\cal M}^2(m^2,m^2)$ 
is the free mass operator and the normalization constant $N$ is 
determined by the condition $F_{\pi}(0)=1$. 

Finally, the pion electromagnetic form factor expressed with the light-front wave 
function for the non-symmetric vertex function, is writing as 
\begin{eqnarray}
F_{\pi}^{+(i)(NSY)}
(q^2) & = & \frac{m^2}{p^+ f^2_\pi} N_c 
\int \frac{d^{2} k_{\perp} d x}
{2(2 \pi)^3 x }  
\biggl[ 
-4 (\frac{f_1}{x p^+})
(x p^+ - p^+)^2 
+ 4 f_1
(x p^+-2 p^+) 
\nonumber \\ 
& & 
+ \ x p^+ q^2 \biggr] 
 \Psi^{*(NSY)}_f(x,k_{\perp}) 
\Psi^{(NSY)}_i(x,k_{\perp}) \theta(x) \theta(1-x). 
\label{form}
\end{eqnarray}

But in the light-front approach, besides the valence 
contribution for the electromagnetic current, the non-valence 
components give another contribution, too~\cite{Pacheco99,Pacheco98,Naus98}. 
The non-valence components contribution is 
calculated in the second interval of the integration~(ii), 
through the "dislocation pole method", developed in reference~\cite{Naus98}. 
The non-valence contributions to the electromagnetic form factor in this case, 
is given:
\begin{eqnarray}
\hspace{-2.0cm} 
F^{+(ii)(NSY)}_{\pi}(q^2) 
 = 
\lim_{\delta^+ \rightarrow 0}  
2 \imath e \frac{m^2 N^2}{ 2 p^+ f^2_\pi} N_c 
\int \frac{
d^{2} k_{\perp} d k^{+}}{2(2 \pi)^4} 
\biggl[
\frac{ Tr~[ {\cal O }^{+}] }
{k^+(p^+-k^+)^2
 (p^{+}-k^+)^2
}  
  \nonumber  \\ 
\hspace{-1.5cm}
\times 
\frac{\theta(p^+-k^+) 
\theta(p^{\prime +} - k^+)} 
{(p^- - \bar{k}^- - \frac{f_2 -\imath \epsilon }{p^+ - k^+})
(p^{-} - \bar{k}^- - \frac{f_3 -\imath \epsilon }{p^{+} - k^+}) 
(p^- - \bar{k}^- - \frac{f_4 -\imath \epsilon }
{p^+ - k^+}) (p^{'-} - \bar{k}^- - 
\frac{f_5 -\imath \epsilon }{p^{'+} - k^+})} 
\biggr]
\nonumber \\
\propto \delta^+ = 0 \ .  
\label{eq5.63}
\end{eqnarray}

As can be seen in equation Eq.~(\ref{eq5.63}),~the electromagnetic form 
factor is directly proportional to $\delta^+$, and that term goes 
to zero with $\delta^+$. Then, the non-valence or the pair term contribution for 
the pion electromagnetic form factor is zero, in the case of non-symmetric vertex for the 
plus component of the electromagnetic current calculated in the Breit frame~\cite{Pacheco99}.

Also, with the minus component of the electromagnetic current,~$J^{-}_{\pi}~(=J^0-J^3$), 
it is possible to extract the pion electromagnetic 
form factor with non-symmetric vertex~Eq.~(\ref{nosymm}). 
But in this case, we have two contributions, one is the valence contribution 
for the wave function 
and the second is the non-valence contribution to the electromagnetic matrix elements of 
the electromagnetic current~\cite{Pacheco99,Pacheco2002,Naus98}. 
The pion electromagnetic form factor for the minus component 
of the electromagnetic current,~$J_{\pi}^{-}$, is related with the Dirac 
matrix by $\gamma^{-}=\gamma^{0}-\gamma^{3}$, as 
known in the light-front approach~\cite{Brodsky98,Harina96}. 
With the non-symmetric vertex,  the minus
component of the electromagnetic current is given by, 
\begin{eqnarray}
 J^{-(NSY)}_\pi &  = &   
e (p+p^{\prime})^{-} F^{-(NSY)}_\pi(q^2) 
\nonumber \\
 & = &  \imath  e^2
\frac{m^2}{f^2_\pi} 
N_c\int \frac{d^4k}{(2\pi)^4} 
Tr \left[ \frac{\psla{k}+m}{k^2-m^2 + \imath \epsilon } 
\gamma^5 \frac{\psla{k}-\psla{p'}+m}{(p'-k)^2-m^2+ \imath \epsilon}
\gamma^{-}   \right. \nonumber \\
& &  \left. \times \frac{\psla{k}-\psla{p}+m} {(p-k)^2-m^2+\imath \epsilon}  
\gamma^5 
\Gamma(k,p^{\prime}) \Gamma(k,p)
\right] .
\label{j-pion}
\end{eqnarray} 
The Dirac trace in equation Eq.~(\ref{j-pion}), for the minus component of 
the electromagnetic current, calculated with the light-front approach, results 
in the following expression: 
\begin{eqnarray}
Tr[ {\cal O }^{-}] & = & \bigl[ -4 k^{-2} k^{+}     
- 4 p^{+} ( 2 k^2_{\perp} + k^+ p^+ + 2 m^2) 
\nonumber \\
& & + k^{-} (4 k_{\perp}^{2} 
+ 8 k^+ p^+ +q^+ + 4 m^2) \bigr].
\label{tracejm}
\end{eqnarray}
In order to calculate the  pair terms contribution for the minus 
component of the electromagnetic current in the 
second interval integration,~($p^+ < k^+ < p^{\prime +}$), the $k^{-}$ 
dependence in the trace is performed and the matrix element of the pair terms are written in 
the equation below:
\begin{eqnarray}
\hspace{-0.9cm}
J^{-(ii)\;(NSY)}=  
\lim_{\delta^+ \rightarrow 0}  2 \imath e \frac{m^2 }{ f^2_\pi} N_c 
\int \frac{ d^{2} k_{\perp} d k^{+}}{2(2 \pi)^4} 
\biggl[ \frac{Tr[ \ {\cal \bar{O} }^{-}] } {k^+ (p^+ - k^+) 
(p^{\prime +} -k^+) 
} \nonumber  \\  
\hspace{-0.90cm}
\times
\frac{ \theta(p^+-k^+)\theta(p^{\prime +} - k^+)}
{ 
(\bar{k^-}-\frac{f_1 - \imath \epsilon}{k^+}) 
(p^- - \bar{k}^- - \frac{f_2 -\imath \epsilon }{p^+ - k^+}) 
(p^- - \bar{k}^- - \frac{f_4 -\imath \epsilon }
{p^+ - k^+}) (p^{'-} - \bar{k}^- - 
\frac{f_5 -\imath \epsilon }{p^{'+} - k^+})
} 
\biggr], 
\label{jmin}
\end{eqnarray}
where $p^{\prime +}=p^+ + \delta^+$ and 
$\bar{k}^{-}=p^- -
\frac{f_3-\imath \epsilon}{p^{\prime+}-k^+}$. 
The pair terms contribution for the minus component of the 
electromagnetic current is obtained with~Eq.~(\ref{jmin}), and the 
Breit frame is recovered in the limit $\delta^+ \rightarrow 0$:
\begin{eqnarray}
J_{\pi}^{-(ii)\ (NSY)}  =  
4 \pi \biggl( \frac{m_{\pi}^2+q^2/4}{p^+} \biggr) 
\int \frac{d^2 k_{\perp}}{2 (2 \pi)^3}
\sum_{i=2}^{5}\frac{ \ln(f_{i})}
{\prod_{j=2,i\neq j}^{5}(-f_i + f_j)}.
\label{jmin2} 
\end{eqnarray}   

The pion electromagnetic form factor with the non-valence contribuition is built 
with the minus 
component of the matrix elements of the electromagnetic current 
calculated in Eq.~(\ref{jmin2}):
\begin{eqnarray}
\hspace{-1.9cm}
F_{\pi}^{-(ii)\ (NSY)}(q^2)  =  
\frac{N^2}{2 p^{-}} \frac{m^2}{f^2_{\pi}} N_c
\biggl( 4 \pi \frac{m_{\pi}^2+qs^2/4}{p^+} \biggr) 
\int \frac{d^2 k_{\perp}}{2 (2 \pi)^3}\sum_{i=2}^{5}\frac{ \ln(f_{i})}
{\prod_{j=2,i\neq j}^{5}(-f_i + f_j)}. 
\end{eqnarray}

The full electromagnetic form factor of the pion, is the sum of the partial 
form factors~$F_{\pi}^{-(i)}$~and 
$F_{\pi}^{-(ii)}$, 
\begin{equation}
F_{\pi}^{-(NSY)}(q^2)=
\left[ 
F_{\pi}^{-(i)(NSY)}(q^2)+
F_{\pi}^{-(ii)(NSY)}(q^2)\right] .
\end{equation}
If the pair terms are not taken into account, 
the rotational symmetry is broken and 
the covariance is lost for the
$J_{\pi}^{-}$ component of the electromagnetic current, 
as can be seen in Fig.~1. After the pair terms or zero modes contribution add in the 
calculation of the electromagnetic form factor with the minus component of the 
electromagnetic current, the following identity is obtained, 
\begin{equation}
F_{\pi}^{-(NSY)}(q^2)=F_{\pi}^{+(NSY)}(q^2)~,
\end{equation}
and the full covariance is restored.

In the next step, it is employed the symmetric 
vertex~$\pi-q\bar{q}$ with the plus component, "+", 
of the electromagnetic current,~Eq.~(\ref{symm}), as utilized in reference~\cite{Pacheco2002}. 

This vertex is symmetric by the exchange of the quadri-momentum
of the quark and the anti-quark. In the light-front coordinates it is written as, 
\begin{eqnarray} 
&  &  \Gamma(k,p) =
{\cal N}  \left[k^+\left(k^{-} -
\frac{k^2_{\perp}+m^2_R -\imath\epsilon}{k^+}
\right)  \right]^{-1}
+ \nonumber \\
& &  {\cal N} \left[(p^+ - k^+)
\left(p^- - k^- - \frac{(p-k)^2_{\perp}+m^2_{R}-\imath\epsilon}
{p^+ - k^+} \right)
\right]^{-1}.
\label{syvertex}
\end{eqnarray}
With the symmetric vertex, the pion valence wave function 
results in the expression
\begin{eqnarray} 
& & \Psi^{(SY)}(x,\vec k_\perp)  =  \\ \nonumber 
& &  
\left[\frac{{\cal N}}
{(1-x)(m^2_{\pi}-{\cal M}^2(m^2, m_R^2))} 
+  
\frac{{\cal N}}
{x(m^2_{\pi}-{\cal M}^2(m^2_R, m^2))}
\right]
\frac{p^+}{m^2_\pi-M^2_{0}}~.
\label{wf2}
\end{eqnarray}
The electromagnetic form factor for the pion valence wave function, 
the expression above, calculated in the Breit frame~$(q^{+}=0)$, is
\begin{eqnarray}
F_\pi^{(SY)}(q^2) & = &  
\frac{m^2 N_c}{p^+f_{\pi}^2}
\int \frac{ d^{2} k_{\perp}}{ 2(2\pi)^3 }
\int_0^{1} \- \- dx
\left[ k_{on}^- p^{+ 2} +   
\frac14 x p^{+} q^2 
\right ]  \nonumber \\
 & & \times 
 \frac{\Psi^{*(SY)}_{f}(x,k_\perp)\Psi^{(SY)}_{i}(x,k_\perp)}{x (1-x)^2},
\end{eqnarray}
where $k_{on}^-=(k_{\perp}^2+m^2)/k^+$ and the normalization
constant ${\cal N}$ is determined 
from the condition $F^{SY}_\pi(0)=1$. 
The pion electromagnetic form factor calculated 
with the symmetric wave function is presented the Fig.~1 
for higher momentum, and in~Fig.~2 for low momentum transfer.  
In both regions, the differences between the symmetric and 
non-symmetric vertex are not so large.

The pion decay constant, measured in the weak leptonic decay, 
is given, with the partial axial current conservation by, 
$P_{\mu} <0|\bar{q} \gamma^\mu \gamma^5 \tau_i q/2|\pi_j> = 
\imath m^2_{\pi}\delta_{ij} $~\cite{Pacheco99,Pacheco2002}, 
and with the the vertex function $\Gamma(k,p)$, is writen by, 
\begin{eqnarray}
 \imath f_{\pi}P^{2}~=~N_{c} \frac{m}{f_{\pi}} \int\dfrac{d^4k}{(2\pi)^{4}}
 Tr \left[\psla{p} \gamma^{5} S(k) \gamma^5 S(k-p)  \right]  \Gamma(k,p).
\label{decay1}
\end{eqnarray}

\newpage

In the case of the symmetric and non-symmetric vertices, 
(see Eqs.~(\ref{nosymm}) and (\ref{symm})),
the expressions for the decay constant are, respectively, 
\begin{eqnarray}
f^{(SY)}_{(\pi)}~=~ \dfrac{m^2 N_c}{f_\pi }
\int \dfrac{d^2k_\perp dx }{4 \pi^3 x (1-x)}
\Psi_{\pi}^{(SY)} \left(x,\vec k_{\perp};m,\vec{0}\right)~,
\end{eqnarray}
and 
\begin{eqnarray}
 f^{(NSY)}_{(\pi)}~=~ \dfrac{m^2N_c }{f_\pi }
 \int\dfrac{d^{2}k_\perp}{4\pi^3}~\dfrac{dx}
 { x }
 \Psi_{\pi}^{(NSY)} \left(k^+,\vec k_{\perp};m,\vec{0}\right)~.
\end{eqnarray}

In the numerical calculations,(see the results section), the obtained values 
of the decay constant
with the expressions above, for both models of light-front calculations, 
do not have significant discrepancies. 

In Next section, the vector meson dominance is presented. 

\section{Vector Meson Dominance} 

In the 1960's, Sakurai~\cite{Sakurai60,Feynman72} proposed the theory of 
{\it Vector Meson Dominance} (VMD); 
a theory of strong interactions with the local gauge invariance,
mediated by vector mesons and basead on the non-Abelian field theory 
of Yang-Mills. However, it is possible to have two lagrangian formulations of the 
vector meson dominance, the first  was introduced by 
Kroll, Lee and Zumino~\cite{Kroll67} and is customary called 
VMD-1. The pion 
electromagnetic form factor calculated with this formulation of the 
vector meson dominance, results in: 
\begin{equation}
F^{VMD1}_{\pi}(q^2)=
\left[ 1-
\frac{q^2}{q^2 - m^2_{\rho}}
\frac{g_{\rho \pi \pi }}{g_{\rho}}  
\right] \; .
\label{vmd1}
\end{equation}

This equation for the electromagnetic form factor satisfies the condition $F_{\pi}(0)=1$, 
indenpendent of assumption about the 
counpling constants,~$g_{\rho \pi \pi}$ and $g_{\rho}$.

In the second formulation of the vector meson dominance, 
the Lagrangian has a photon mass term, and the photon 
propagador has a non-zero mass; that version is usually called~VMD-2.
With this second formulation of the vector meson dominance, 
the pion electromagnetic form factor is written:
\begin{equation}
F^{VMD2}_{\pi}(q^2)=
\left[-
\frac{m^2_{\rho}}{q^2 - m^2_{\rho}}
\frac{g_{\rho \pi \pi }}{g_{\rho}}  
\right] \; .
\label{vmd2}
\end{equation}
In the equation above, 
it is necessary that, the condition $F_{\pi}(0)=1$ is satisfied, only if the 
 universality limit is taken into account, or, translate in the following 
equality,~$g_{\rho \pi \pi} ~=~g_{\rho}$. 
In the universality limite, like advocate by J.~Sakurai, the two formulations of the 
vector meson dominance are equivalent. 

For the present work, in Eq.~(\ref{vmd1}) and Eq.~(\ref{vmd2}), 
the rho meson mass utilized is the experimental value, 
$m_\rho=0.767$~GeV, and, from the universality, $g_{\rho \pi \pi}=g_{\rho}$, 
the results at zero momentum for both equations satisfy $F_{\pi}(0)=1$.

In the present case here,  only the 
lightest vector resonance rho meson is taken account in the 
monopole model of the VMD as can be seen in Eq.~(\ref{vmd1}) or 
Eq.(\ref{vmd2}). 
The vector meson dominance works quite well in the
timelike region below the $\pi\pi$ threshold.
At low energies for the space-like region, 
the vector meson dominance model 
gives a reasonable 
description for the 
pion electromagnetic form factor. For more details and results about 
the vector meson dominance, see references~\cite{Krein93,Pacheco2005,Pacheco20052}. 

\section{Results}   

The pion electromagnetic form factor, 
presented consistently with previous and later works, are extended at higher momentum 
transfer region, 
running $Q^2=-q^2$ up to 20~(GeV/c)$^2$. Here, the models of the 
\linebreak $\pi-q\bar{q}$ verticies, 
i.e, non-symmetric and symmetric verticies~\cite{Pacheco99,Pacheco2002} 
are compared with the vector meson dominance (VMD), and 
shown in Figs.~1,~2 and 3,  for low and higher low momentum transfer.  
The pion electromagnetic radius, is calculated 
with the derivative of the electromagnetic form factor for the 
pion,
\linebreak ~$<r^2>=-6 dF(q^2)/dq^2_{|q^2 \simeq 0}$, 
for both model of the verticies presented here.

In the case of the non-symmetric vertex, 
the pion radius is utilized to fix the parameters of the
model. The parameters are the quark mass 
$m_q=0.220$~GeV and the regulator mass $m_R=1.0$~GeV. The pion mass utilized is the experimental
value, $m_{\pi}=0.140$~GeV. The experimental radius of the pion is
$r_{exp}=0.672\pm0.02$~fm~\cite{Amendolia84,PDG}. 

Using the pion decay constant calculation in the non-symmetric 
vertex model and with the parameters above,  the 
pion decay constant obtained is~$f_{\pi}=92.13$~MeV, 
which is close to the experimental value,~$f_{\pi} \simeq 92.10$~\cite{PDG}.

In the case of the symmetric vertex, the parameters are the quark mass~$m_{q}=0.220$~GeV, 
the regulator mass~$m_{R}=0.60$~GeV and the experimental mass
of the pion, $m_{\pi}= 0.140$~GeV. 

Our choice for the regulator mass, fits the pion
decay constant, $f_{\pi}^{exp}=92.1$~MeV, for the symmetric vertex,
quite well compared with the experimental data~\cite{PDG}. 

Both light-front models, with symmetric and non-symmetric vertex, 
have good agreement with the experimental data 
at low energy, however, 
some differences are noticiable in the region $Q^2 \geq$~1.0 (GeV/c)$^2$ (see the fig.~1). 
The experimental data collected from reference~\cite{Frascati2001} are described 
well up to 10~(GeV/c)$^2$~with both the symmetric and non-symmetric vertex functions. 
For the minus component of the electromagnetic current,~$J^-$,~the pair terms or non-valence components 
of the electromagnetic current contributions, 
are essential to obtain the full covariant pion electromagnetic form factor, 
and to the end to respect the covariance.

\begin{center}
\begin{table}[tbh]
 \caption{ Results for the low-energy electromagnetic 
 $\pi$-meson observables with light-front and another models .} 
 \vspace{0.20cm}
 \item[]
\begin{tabular}{l|l|l|l}
\hline 
 Model & $r_{\pi}$ (fm)      & $f_{\pi}$ (MeV) &  $r_{\pi}.f_{\pi}$ \\
\hline
Sym. (LF)                    &~0.740  &~92.40  &~0.346 \\
Non-Sym.(LF)                 &~0.679  &~93.10 &~0.320 \\
LFBS Model~\cite{Choi2001}   &~0.651  &~91.91 &~0.304 \\ 
Dyson-Sch.~\cite{Maris98}    &~0.550  &~92.0  &~0.256\\
KLZ Model~\cite{Loewe2007}   &~0.631  & ~~-   & ~~- \\ 
Exp.~\cite{PDG}              &~0.672$\pm$0.02 &~92.1 &~0.314$\pm$0.010 \\  
\hline
\end{tabular}
 \label{table1}
\end{table}
 \end{center}
 
   \begin{figure}
       \centerline{\includegraphics[width=14 cm,height=14 cm]
                                   {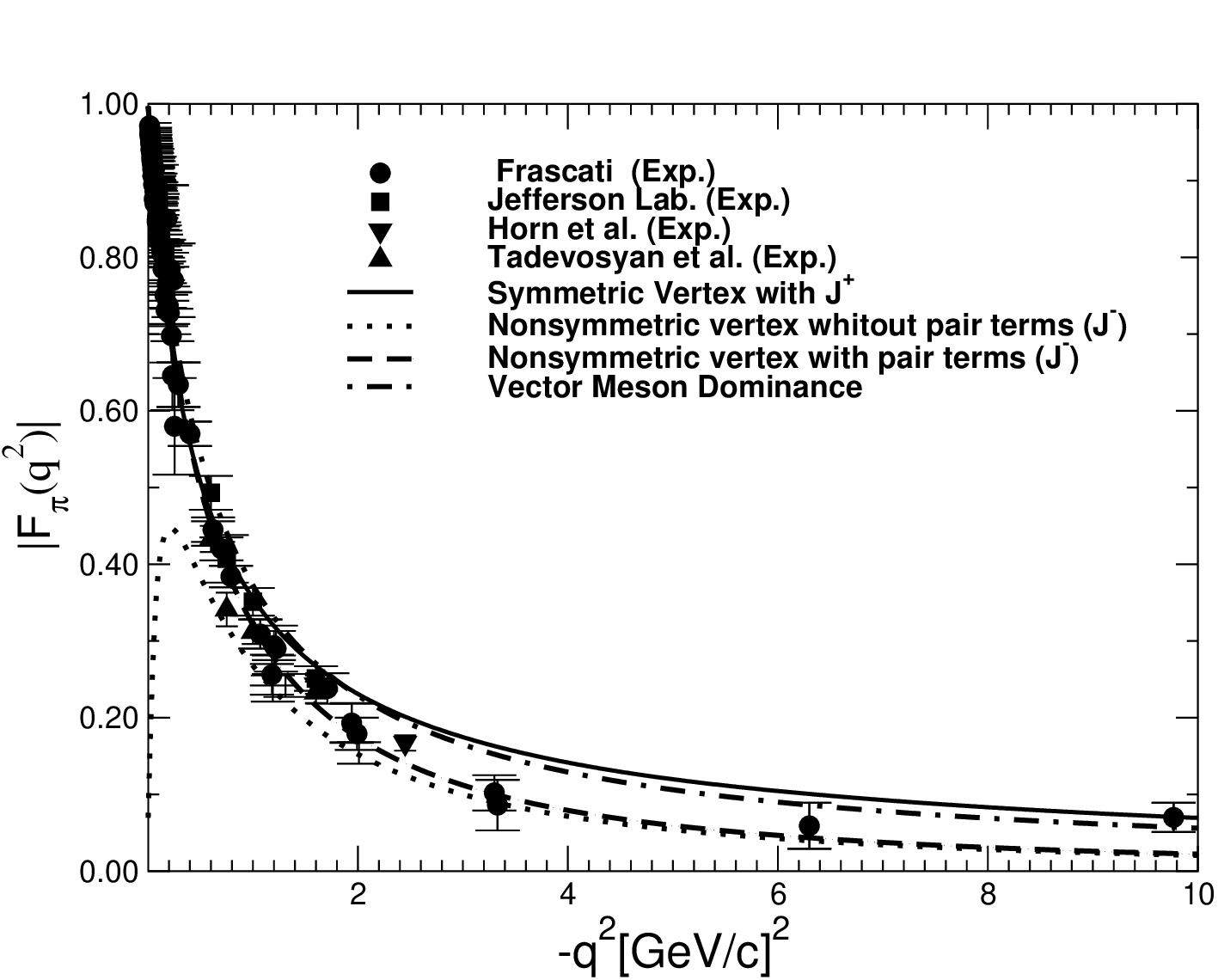}}  
\caption{Pion electromagnetic form factor calculated with light-front 
constituent quark model, for the  plus and minus components of 
electromagnetic current, compared with experimental data and vector meson 
dominance. Data are from~\cite{Volmer2001,Frascati2001,Horn06,Tadevosyan2006}. 
Solid line is the full covariant form factor with $J^+_{\pi}$
 (symmetric vertex for the $\pi-q\bar{q}$).
 The dashed line is line the form factor 
 with $J_{\pi}^{-}$ plus pair terms
 contribution, and the dotted line is the pion form factor without  
 the pair terms contribution with the 
 minus component of the electromagnetic current, where both curves are with the nonsymmetric vertex. 
 After added the non-valence contribuition, the 
 pion electromagnetic form factor calculated with the plus or minus 
 compoenent of the electromagnetic current give the same results for the nonsymmetric 
 vertex.}
\label{Fig.1}
\end{figure}

Constituent quark models formulated with the light-front approach presented here, 
give a good agreement with the  
experimental data~\cite{Amendolia84,Amendolia86,Frascati2001,Volmer2001,Horn06,Tadevosyan2006}.  
The ratios between the electromagnetic current 
in the light-front and the electromagnetic current calculated in the 
instant form, are given by the following equations, 
\begin{eqnarray}  
Ra^{I} & = & \frac{J^{+}_{LF}}{J^{+}_{Cov}}, 
\hspace{2.7cm} Ra^{II} = \frac{J^{-}_{LF}}{J^{-}_{Cov}}, 
\nonumber \\
Ra^{III} & = & \frac{J^{-}_{LF}+J_{LF}^{- (Pair)}}{J^{-}_{Cov}}
, 
 \ \ \ \ \ Ra^{IV}  =  \frac{J^{-}_{LF}}{J^{+}_{Cov}}, \nonumber \\
Ra^{V} & = & \frac{J^{-}_{LF}+J_{LF}^{- (Pair)}}{J^{+}_{Cov}}, 
\label{raza}
\end{eqnarray}
where the non-symmetric vertex is utilized according 
to~Eq.~(\ref{nosymm}).

In~Eq.~(\ref{raza}),~above , $Ra^{I}$ is the 
plus component of the electromagnetic current calculated 
in the light-front divided that of 
the instant form formalism, since the pair terms do not give contribution for the plus 
component of the electromagnetic current, so the ratio $Ra^I$ is 
constant~(see Fig.~\ref{Fig.4}). The second ratio, $Ra^{II}$, is the minus component 
of the electromagnetic current, $J^{-}$, 
calculated with the light-front formalism and divided by the electromagnetic current 
calculated in the instant form. In $Ra^{III}$ ratio, the pair terms 
contribution to the 
electromagnetic current is included, so the covariance is restorated.

The ratios $Ra^{IV}$ and $Ra^{V}$ are the "minus" components of the 
electromagnetic current without and with the pair terms 
contribution,~respectively,~divided by the "plus" component of the electromagnetic 
current calculated in the instant form formalism.

\begin{figure}
\centerline{\includegraphics[width=14 cm,height=14 cm]
                                   {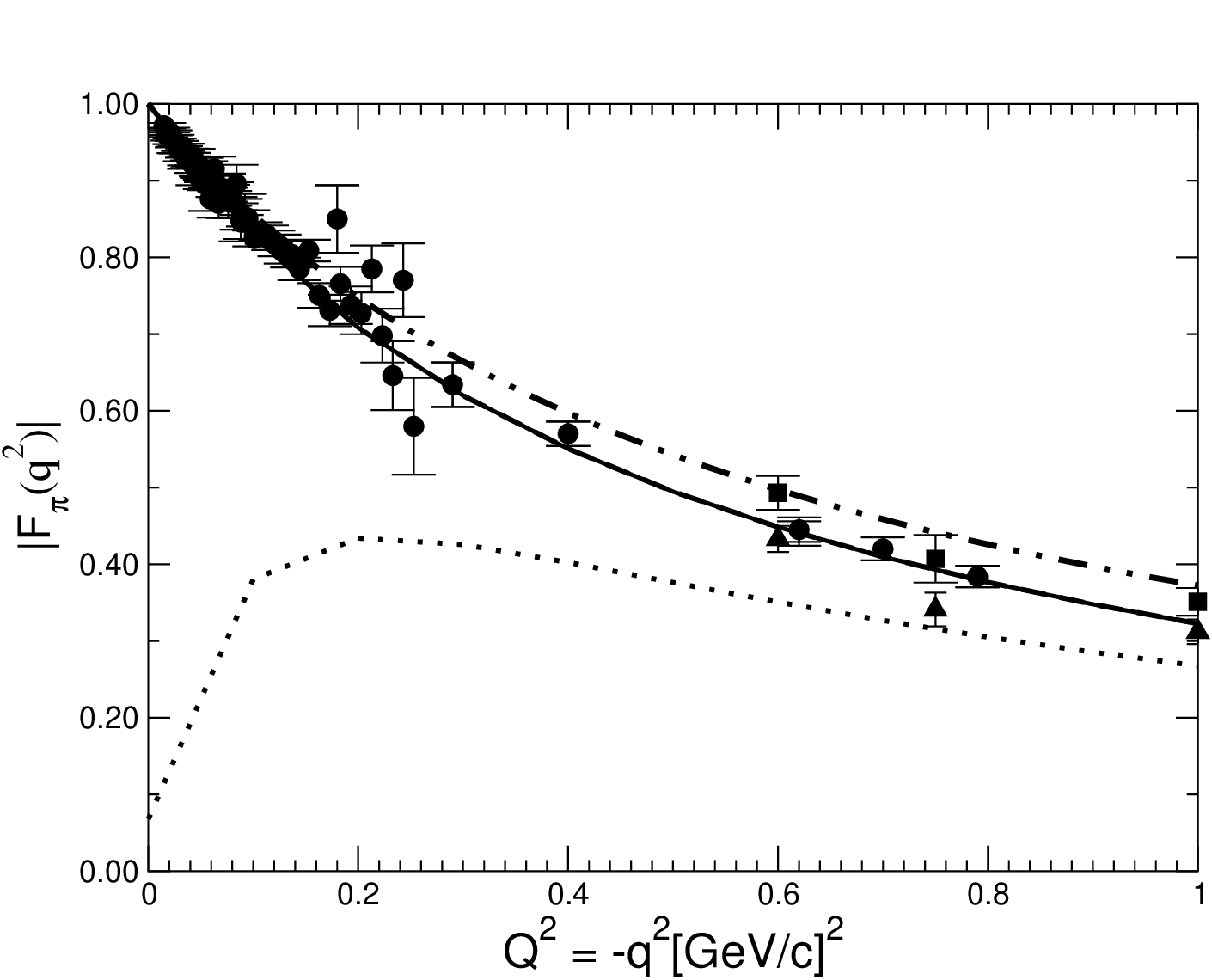}} 
\caption{Pion electromagnetic 
form factor for small $Q^2$.~Labels are the 
same as those in Fig.~\ref{Fig.1}.
 Experimetal data are from 
 Ref.~\cite{Amendolia86,Frascati2001,Volmer2001}.}
\label{Fig.2}
\end{figure}

   \begin{figure}
       \centerline{\includegraphics[width=14 cm,height=14 cm]
                                   {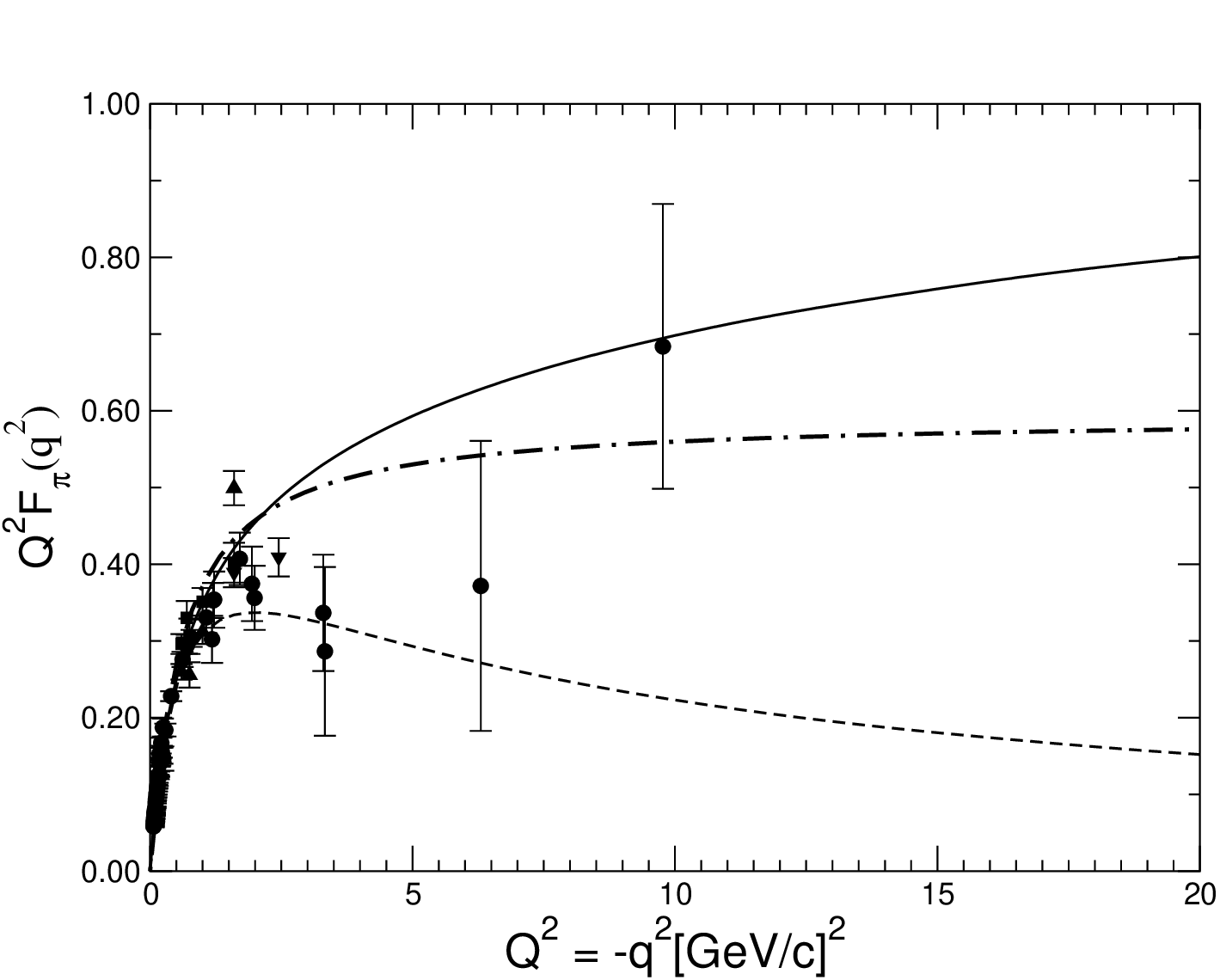}}
\caption{Pion electromagnetic 
form factor for higher $Q^2$. Labels are the 
same as those in Fig.~1.}
\label{Fig.3}
               \end{figure}

As can be seen in Fig.~\ref{Fig.4}, 
the rotational symmetry in the 
light-front formalism is broken, 
because the pair terms or non-valence contribution for the 
electromagnetic current is not taken into account properly. 
The restoration of the symmetry breaking is obtained 
by adding the 
pair terms contribution to the minus component of the electromagnetic 
current calculated in the light-front.

Experimental data for $Q^2 \gtrsim 1.5~$(GeV/c)$^2$ for the pion electromagnetic form factor
~(see the Fig.~\ref{Fig.1}), is not precise in order to make 
a decision satisfactorily among the phenomenological models to 
select a best description for the 
pion elastic electromagnetic form factor nor in the end, the correct pion wave function.

We define the following equations, in order to compare 
the magnitude of the breaking of the rotational symmetry 
for the pion electromagnetic form factor calculation   
with the light-front models, and vector meson dominance model, 
and, also with the covariant calculations; 
the strategy, is that we try to amplify the differences among theoretical 
models and experimental data:
\begin{eqnarray}
\Delta_1   & = & \left[   
q^2 F^{(VMD)}_{\pi}(q^2) - q^2 F^{+(NSY)}_{\pi}(q^2) \right] \; ,  
 \nonumber \\ 
\Delta_2 & = & \left[ 
  q^2 F^{(COV)}_{\pi}(q^2) - q^2 F^{-(i)(NSY)}_{\pi}(q^2) \right] \; ,
\nonumber \\
\Delta_3 & = & \left[ 
  q^2 F^{(COV)}_{\pi}(q^2) - q^2 F^{-(i+ii)(NSY)}_{\pi}(q^2) \right] \; ,
\nonumber \\
\Delta_4 & = & \left[ 
  q^2 F^{(VMD)}_{\pi}(q^2) - q^2 F^{(exp)}_{\pi}(q^2) \right]\; .
\label{diffe2}
\end{eqnarray}

The results of the calculations above are shown in the 
Figs.~\ref{Fig.5} and \ref{Fig.6} at 
low and higher momentum transfer for the models presented here. 
The results in Fig.~\ref{Fig.5}, 
confirm the validity of the vector 
meson dominance model at very low momentum transfer ($Q^2\leq 0.5$~(GeV/c)$^2$).

But, for $Q^2>~0.5~$(GeV/c)$^2$ (see Fig.~\ref{Fig.6}), the discrepancies between the 
vector meson dominance model, the light-front models and 
experimental data are more emphasized.  
In the case of $\Delta_3$,~(see the definition above in the text), 
the covariance is respected exactly, because the difference is zero in 
the interval integration sum,  [(i)+(ii)], for the $J^{-}$ component of the 
electromagnetic current.

The electromagnetic form factor for the 
pion calculated with the matrix elements of the electromagnetic current gives the same results as the
electromagnetic form factor of the pion calculated with usual covariant 
quantum field theory~\cite{Zuber}.  
   \begin{figure}
       \centerline{\includegraphics[width=14 cm,height=14 cm]
                                   {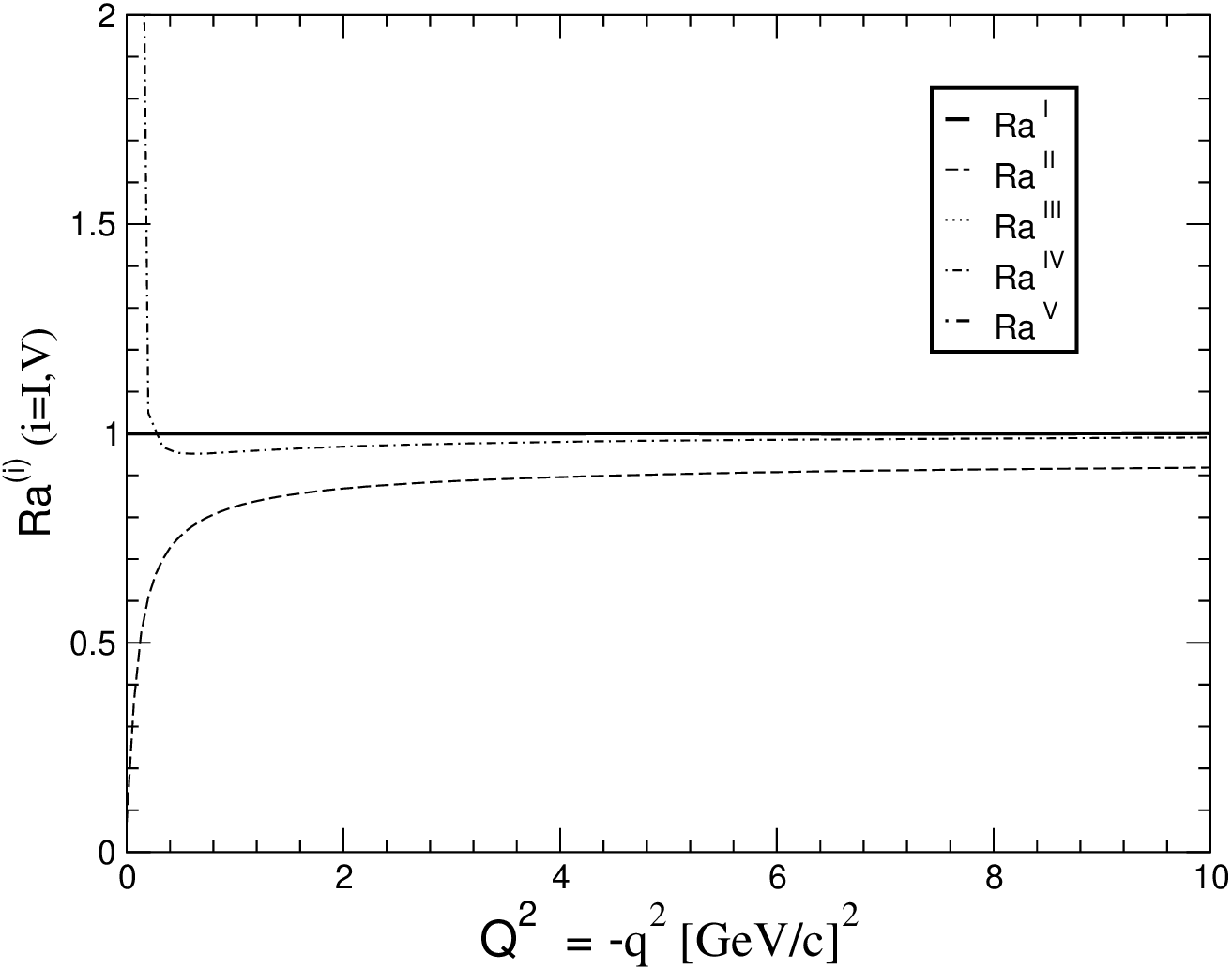}}
\caption{Pion electromagnetic 
current ratios, see Eq.(\ref{raza}) in the text.}
\label{Fig.4}
               \end{figure}

   \begin{figure}
       \centerline{\includegraphics[width=14 cm,height=14 cm]
                                   {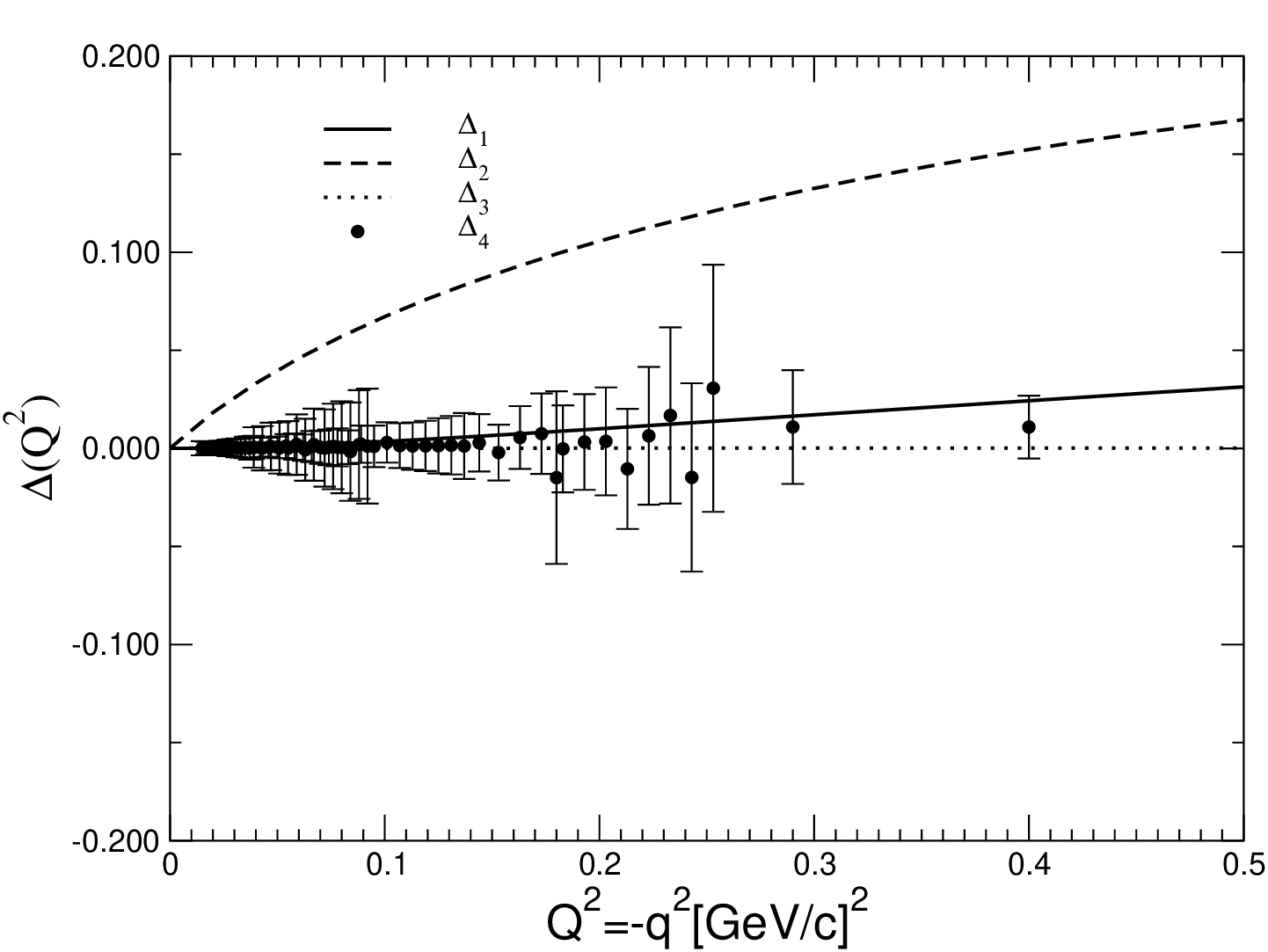}}
\caption{Figure labels are the same as those in 
Eq.(\ref{diffe2}). The range for the 
momentum transfer given here is up to 0.5~(GeV/c)$^2$.
}
\label{Fig.5}

               \end{figure}

   \begin{figure}[htb]
       \centerline{\includegraphics[width=14 cm,height=14 cm]
                                   {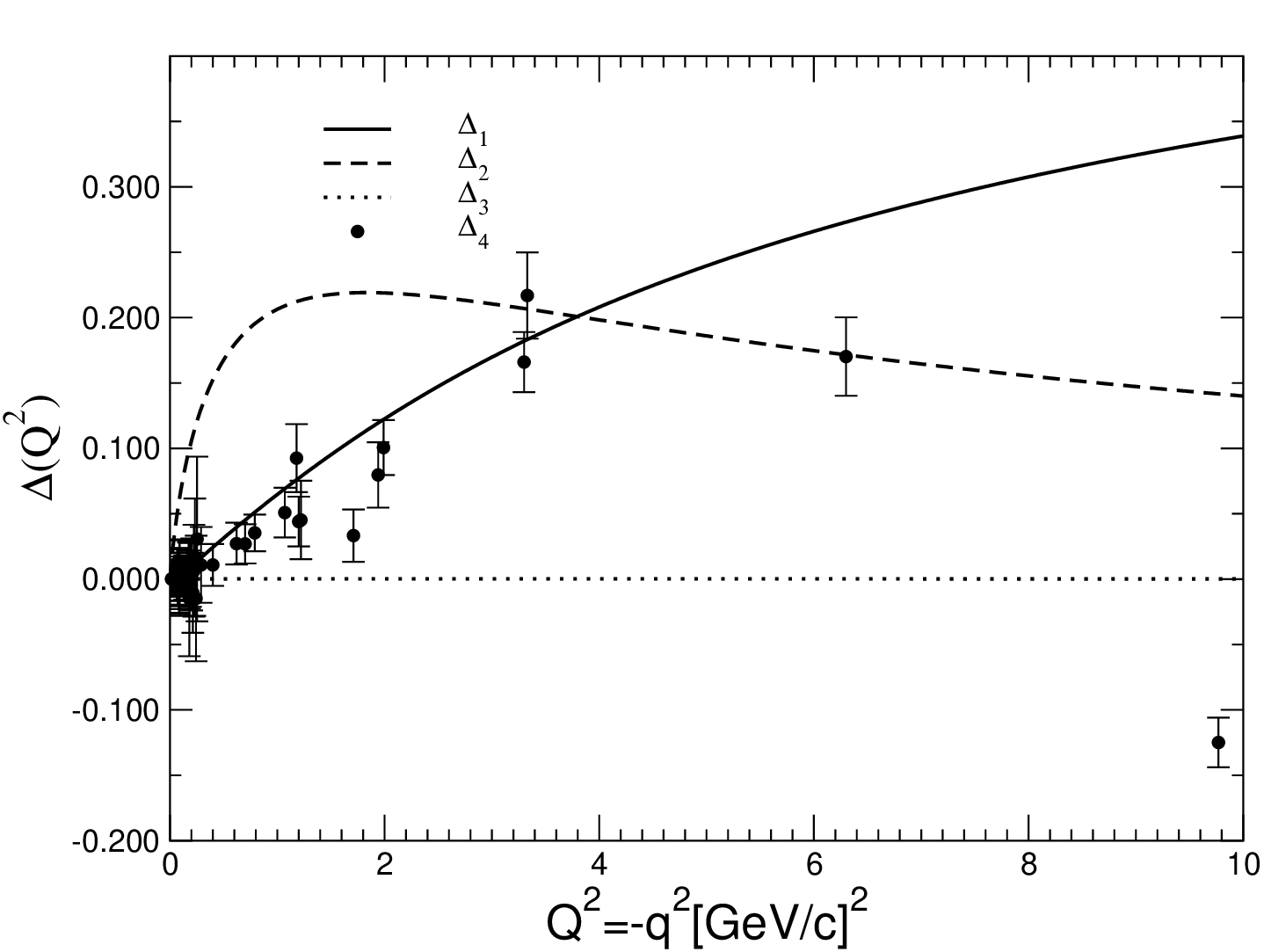}}
\caption{Figure labels are the same as those in figure 4
(see also Eq.(\ref{diffe2}), here the momentum transfer is 
up 10~(GeV/c)$^2$. }
\label{Fig.6}
\end{figure}

For the higher momentum transfer, 
the asymptotic behavior for the wave function of the non-symmetric vertex 
model produce  $q^2F_{\pi}$15~(GeV/c)$^2$ $\approx 0.18~$(GeV/c)$^2$. 
That result is compared  with the  leading-order-perturbative QCD, 
 $Q^2F_{\pi}(Q^2)\approx 0.15$~(GeV/c)$^2$, 
 for $\alpha_s(Q^2=10$~(GeV/c)$^2$) $\approx 0.3$ and with 
Dyson-Schwinger approach,~$Q^2F_{\pi}(Q^2)\approx 0.12-0.19~$(GeV/c)$^2$, 
for momentum transfer 
between~$Q^2~\approx ~10-15~$(GeV/c)$^2$~\cite{Maris98}.

\section{Conclusions}

In the present work, the electromagnetic form factor of the pion 
was investigated in the range $0<Q^2<20~$(GeV/c)$^2$ 
with light-front constituent quark model. 
The light-front formalism is known nowadays as a natural way to describe 
the systems with relativistic bound state, like the pion. 
 With this approach it is possible to calculate the electromagnetic 
form factors in a most suitable way.

However, problems related with the broken of the 
rotational symmetry in the light-front formalism are important 
and the pair terms or no-valence terms contribution for the covariance restoration in 
higher energies is also necessary to be taken care of~\cite{Pacheco99,Pacheco97}.

After adding the pair  terms, 
or non-valence components in the 
matrix elements of the electromagnetic current, the 
covariance is completely restored, and it doesn't matter 
which component of the electromagnetic current,~$J^+$~or~$J^-$,
~is utilized in order 
to extract the pion form factor with the light-front 
approach, as can be seen in Figs.~\ref{Fig.1},~\ref{Fig.2} and \ref{Fig.3}.

In terms of the electromagnetic current, the numerical results in~Fig.\ref{Fig.4}, 
show the importance of the non-valence components for the electromagnetic current 
and the dependence which component of the currents utlilized at low 
momentum transfer, the inclussion of the non-valence components of the electromagnetic current 
is essential for the minus component of the electromagnetic current, 
to give the full covariance. 

In Eq.(\ref{raza}) the ratios $Ra^{I}, Ra^{III}$ and $Ra^{V}$, produce 
constant values, but the ratios 
$Ra^{II}$ and $Ra^{IV}$ are not, because the non-valence components of the 
electromagnetic current is not included in the light-front approach 
calculation~(see Fig~\ref{Fig.4}).

The comparison between the light-front models 
for the vertex $\pi-q\bar{q}$ with other hadronic models for the 
pion electromagnetic form-factor has a good agreement between them, 
however, some differences arise between these models when energies are 
in the higher region,~$Q^2 \gtrsim 2$~(GeV/c)$^2$. 

With Eq.(\ref{diffe2}), the diferences between the models analyzed 
in the present work are clear, for lower and higher momentum transfer, because 
the set of equations increase the possible differences among the models 
presented here. 

Since the pion electromagnetic form-factor is sensitive 
to the model utilized, it is important to compare 
different models including new experimental data, and 
to extract new information about the sub-hadronic 
structure of the pion bound state. 

The light-front approach is a good 
framework to study the pion electromagnetic form factor.
However, the inclusion of the non-valence components of the electromagnetic 
current is essential for both low and higher momentum transfer.

To conclude, the light-front formalism and the vertex models for 
$\pi-q\bar{q}$ utilized in the 
present work with symmetric and non-symmetric vertices, 
can describe the new experimental data 
for the pion electromagnetic form 
factor with very good agreement.  In the next step, the calculations for the vector mesons, 
like $\rho$-meson and vector kaon, are in progress in order to 
compare the light-front constituent models with the other models.

\section*{Ackonowlegments}

This work was supported by 
the Brazilian agencies FAPESP, (Funda\c{c}\~ao de 
Amparo \`a Pesquisa do Estado de S\~ao Paulo), 
CNPq (Conselho Nacional de Desenvolvimento Cient\'\i fico 
e Tecnol\'ogico) and 
CAPES,~(Coordena\c{c}\~ao de Aperfei\c{c}oamento de Pessoal de N\'\i vel Superior). 
R\^omulo Moita, also, thanks the {\it Education Secretary} of the states Piau\'i and Maranh\~ao, Brazil, 
for financial support.

"The authors  declares 
that there is no conflict of interest regarding the publication of this paper."


\end{document}